\documentstyle[latex8]{article}

\newtheorem{ithm}{Theorem}

\newtheorem{lemm}{Lemma}

\newtheorem{corol}{Corollary}

\newtheorem{defin}{Definition}
\newtheorem{deff}{Definition}
\newtheorem{prop}{Proposition}

\newcommand {\bbox} {\vrule height7pt width4pt depth1pt}

\begin{document}
\title{Limitations of Noisy Reversible Computation} 
\author{ Dorit Aharonov
\thanks{Department of Computer Science, Hebrew University}
\and Michael Ben-Or
\thanks{Department of Computer Science, Hebrew University}
\and Russell Impagliazzo 
\thanks{Dept. of Computer Science, UCSD}
\and Noam Nisan 
\thanks{Department of Computer Science, Hebrew University}
}

\maketitle        

\begin{abstract}
In this paper we study noisy reversible circuits.  Noisy computation
and reversible computation have been studied separately, and it is known
that they are equivalent in power to unrestricted computation.  We study
the case where both noise and reversibility are combined and show
that the combined model is weaker than unrestricted computation.  

We consider the model of reversible computation with noise, where
the value of each wire in the circuit is flipped with some
fixed probability $1/2>p>0$ each time step, and all the inputs to the circuit
are present in time 0.  We prove that any noisy reversible 
circuit must have size exponential in its depth in order to compute 
a function with high probability.  This is tight as we show that any 
(not necessarily reversible or noise-resistant) circuit can be 
converted into a reversible one that is noise-resistant with a
blow up in size which is exponential in the depth.  This
establishes that noisy reversible computation has the
power of the complexity class $NC^1$. 

We extend the upper bound to quantum circuits, and  prove that
 any noisy quantum 
circuit must have size exponential in its depth in order to compute 
a function with high probability.
This high-light the fact that current error-correction schemes for 
 quantum computation require constant inputs throughout the computation 
(and not just at time 0), and shows that this is unavoidable.
As for the lower bound, we show   
 that quasi-polynomial noisy quantum circuits are at least powerful 
as quantum circuits with logarithmic depth (or $QNC^1$). 
Making these bounds tight is left open in the quantum  case.
\end{abstract}

\section{Introduction}
In this paper we study noisy reversible circuits.  Noisy computation
and reversible computation have been studied separately, and it is known
that they are equivalent in power to unrestricted computation.  We study
the case where both noise and reversibility are combined and show
that the combined model is weaker than unrestricted computation.  

The model of reversible noisy computation seems natural by itself, especially
when viewed as a model of physical computation.  It is also
motivated by the current surge of interest in noise in quantum computation,
which generalize reversible computation.  Indeed, we extend our lower bounds
to the quantum case.

\subsection*{Reversible Circuits}

The subject of reversible computation,  was first  raised
with connection to the question of how much
 energy is required 
to perform a computation.  
 In this paper we do not
wish to argue about when and in what ways this reversibility condition
is indeed a true requirement, but rather limit ourselves to study models
in which this requirement holds. The reader can consult for example
  Landauer\cite{Lan61} and  Bennett\cite{Ben73,Ben82,Ben89} 
for more discussion.

Our model of reversible computation will be
boolean circuits which may use only reversible gates.

\begin{defin}
A function $g: \{0,1\}^k \rightarrow \{0,1\}^k$ is called reversible
if it is 1-1 (thus, a permutation).  A gate with $k$ inputs and $k$ outputs
is called reversible if it computes a reversible function.
\end{defin}

Our circuits will be composed of reversible gates which belong to some
fixed set of gates (our computation basis).  As is usual in boolean circuits
the exact choice of a base is not important as long as it is finite and
``universal''.  An example of a reversible gate which, by itself, is
a universal family is the 3-input, 3-output Toffoli gate.  
In order to keep reversibility, we do not allow wires in the circuit
to ``split'', i.e. the fanout of each output wire (or input bit) in the circuit
is always 1.  It follows that a reversible circuit has $N$ inputs and
$N$ outputs, for some $N$, and that it computes a reversible function.

In order to compute non-reversible functions by reversible circuits
we apply the following convention.

\begin{defin}
We say that a reversible function $F:\{0,1\}^N \rightarrow \{0,1\}^N$
implements a boolean function $f:\{0,1\}^n \rightarrow \{0,1\}$ if
for every $x \in \{0,1\}^n$, the first output bit of $F(x \vec{0})$ is
$f(x)$.  Here $\vec{0}$ means padding $x$ by $N-n$ 0-bits in order to
get an $N$-bit input for $F$.
\end{defin}

The most basic simulation result of general circuits by reversible
circuits states:

\begin{prop}
If a boolean function $f:\{0,1\}^n \rightarrow \{0,1\}$ can be computed
by a circuit of size $s$ and depth $d$, then it can be implemented by
reversible circuits of size $O(s)$ and depth $O(d)$.
\end{prop}

More advanced simulation results are also known, e.g. that the output
on input $x0$ can be forced to be $x \circ f(x) \circ \vec{0}$, and not just
an arbitrary string starting with $f(x)$.

\subsection*{Noisy Circuits}

Normal boolean circuits are very sensitive to ``hardware failures'':
if even a single gate or single wire malfunctions then the computation
may be completely wrong.  If one worries about the physical possibility
of such failures, then it is desirable to design circuits that are 
more resilient.  Much work has been done on this topic, starting from
Von-Neumann \cite{Von}.

The usual models assume that each gate in the circuit
can fail with some fixed probability $p>0$, in which case its value
is flipped or controlled by an adversary.
Most upper bounds (as well as ours) 
work even for the case of an adversary,
while most lower bounds (as well as ours) 
work even for the simpler case of random flips.  
The aim is to construct circuits that still compute, with high probability,
 the desired function, even when they are noisy.
  The probability of error achieved by 
the circuit (due to the noise) must be at most some fixed 
constant $\epsilon<1/2$.  The exact values of $p$ and $\epsilon$
turn out not to matter beyond constant factors as long as $p$ is 
less than some threshold $p_0$
(which depends on the computational basis), and $\epsilon$ is
at least some threshold $\epsilon_0$ (which depends on 
$p$ as well as on the computational
basis). 
 We say that such a circuit computes
the function in a noise-resistant way.  The basic simulation results
regarding noisy circuits state that any circuit of size $s$ and depth $d$
can be converted into a noise-resistant one of size $poly(s)$ 
and depth $O(d)$ which computes the same function.(For lower bounds on the blow-up in 
depth and size see \cite{Pip,Ann}).

We will be considering a slightly different model in which the errors (noise)
are not on the gates, but rather on the wires.  We assume that each
wire flips its value (or allows an adversary to control its value)
with probability $p$, each ``time unit''.  
This means that we view the depth of a gate in the circuit as
corresponding to the ``time'' in which this gate has its latest
input available (its output will be available one time unit later).
A wire that connects an output of gate at depth $d$ to an input of
a gate of depth $d'$, will thus have $d'-d$ time units in which its value
may be flipped (or controlled by an adversary), with probability $p$ each
time unit.  We call such circuits noisy circuits.

This model does seem reasonable as a model of many scenarios of
noisy computations.  An obvious example is a cellular automata whose
state progresses in time, and it does seem that each cell can 
get corrupted each time unit.  We invite the reader to consider her
favorite
physical computation device and see whether its model of errors agrees
with ours.
In any case, it is easy to see that
this model is equivalent in power to
the one with noise on the gates, and thus is as powerful
as non-noisy circuits\cite{Von}.  

\subsection*{Noisy Reversible Circuits}

The model under study in this paper is the model of reversible
circuits, as defined above, when the wires are subject to noise,
also as defined above.  The combination of these two issues
has not been, as far as we know, studied formally before, though it might seem
natural that given that all operations are reversible, the effect of noise can 
not be corrected. Indeed, it turns out
that this combination is more problematic, in terms of computation
power, than each of the elements alone.

We wish to emphasize that this ``problematic'' behavior appears
only with the combination of the
definitions we consider -- which we feel are the
interesting ones, in many cases.  Specifically, it is not difficult
to see that each of the following variant definitions
of ``noisy reversible''
circuits turns out to be equivalent in power to normal circuits.

\begin{itemize}
\item The noise is on the gates instead of on the wires.
\item The noise on each wire is constant instead of being constant per
time unit.
\item The inputs to the circuit can be connected at an arbitrary 
``level'' of the circuit (corresponding to at an arbitrary time), as
opposed to only at ``time 0''.
\item Constant inputs can be connected at arbitrary levels.
\item The reversible circuits may contain 1-to-1 functions with
a different number of input and output bits.
\end{itemize}

We first show that noisy reversible circuits can simulate general
ones, although with a price which is exponential in the depth.

\begin{ithm}\label{Upper}
If a boolean function $f$ can be computed by a circuit
of size $s$ and depth $d$, then $f$ can be computed by a reversible
noisy circuit of size $O(s \cdot 2^{O(d)})$ and depth $O(d)$.
\end{ithm}

Our main theorem shows that this exponential blowup in depth
is un-avoidable.  It turns out that noisy reversible circuits
must have size which is exponential in the depth in order to do 
anything useful.

\begin{defin}
We say that a noisy reversible circuit is worthless if on every fixed input,
its first output bit takes both values ($0$ and $1$) with probability
of at least $49/100$ each.
\end{defin}

This means that a worthless circuit simply outputs random noise
on its first output bit, whatever the input is.

\begin{ithm}\label{Lower}
For any noisy reversible circuit
of size $s$ and depth $d$ which is not worthless,
$s= 2^{\Omega(d)}$.
\end{ithm}

This give a full characterization of polynomial size noisy reversible
circuits.

\begin{corol}
Polynomial size noisy reversible circuits have exactly the power
of (non-uniform) $NC^1$.
\end{corol}

Note that for the lower bounds
 we do not assume anything on the fan-in of the gates:
In fact, the lower bound still hold even if the gates may operate on 
all the qubits together.
\subsection*{Quantum Circuits}
We extend the upper bound, and a weaker version of the lower bound, to 
quantum circuits, which are  the quantum generalization of 
reversible circuits.
We will be using the model of quantum circuits with mixed states, suggested in
\cite{AN96}.

The computation is performed by letting a system of $n$ quantum bits,
or ``qubits'', develop in time. The state of these $n$ qubits is a vector in 
the complex Hilbert Space $\cal{H}$$^n_2$,
generated by the vectors $|0>,|1>,...|2^{n-1}>$,
where the numbers  $i$ are written in binary representation. 
This vector space is viewed as a tensor product of $n$ copies of $\cal{H}$$_2$,
each corresponding to one of the qubits.
The initial state is one of the basic state which corresponds to the input string.
This state develops in time according to the gates in the circuit.
A quantum gate of order $k$ is a unitary matrix operating in the Hilbert space 
$\cal{H}$$^k_2$ of $k$ qubits.  
Our circuits will be composed of quantum gates which belong to some
fixed set of gates (our computation basis).  As is usual in boolean circuits
the exact choice of a base is not important as long as it is finite and
``universal''\cite{BV93,DiV2}.
Keeping the number of qubits constant in time, we do not allow wires in the circuit
to ``split'', i.e. the fanout of each output wire (or input bit) in the circuit
is always 1.  It follows that a quantum circuit has $N$ inputs and
$N$ outputs, for some $N$.
The function that the quantum
 circuit computes is defined as the result of a ``measurement'' of the first qubit:
i.e. some kind of projection of the final
state on a subspace of $\cal{H}$$^n_2$.
In the model of quantum circuits with mixed states, we allow the $n$ qubits
to be in some probability distribution over vectors in the Hilbert space,
and such a general (mixed) state is best  described by the physical notion of
density matrices. 

\subsection*{Noisy Quantum Circuits}

As in the classical case, we consider the model of noise on the wires.
  We assume that each wire (qubit)
 allows an adversary to control its ``value''
with probability $p$, each ``time unit''.  
The definition of Quantum noise is more subtle than that of classical noise 
since the ``value'' of a qubit is not always defined.
Instead, what we mean by ``controlling the qubit'' is the following operation:
An arbitrary operation on the ``controlled'' qubit and the state of the environment,
represented by $m$ qubits in some arbitrary state, is applied,
after which the state of the environment is averaged upon, to give 
 the (reduced) density matrix to the
 $n$ qubits of the circuit.
This type of damage on a qubit occurs with probability $p$ each time step for
each qubit. The computation is composed alternately of noise steps
 and computation steps.
  We call such quantum circuits noisy quantum circuits.

We first show the quantum analog of theorem \ref{Upper}, i.e.
 that noisy quantum circuits
must have size which is exponential in the depth in order to do 
anything useful.

\begin{ithm}\label{QLower}
For any noisy quantum circuit
of size $s$ and depth $d$ which is not worthless,
$s= 2^{\Omega(d)}$.
\end{ithm}

Where, as for reversible circuits,
we say that a noisy quantum circuit is worthless if on every fixed input,
its first output bit takes both values ($0$ and $1$) with probability
of at least $49/100$ each.

We next give a lower bound on the power of noisy quantum circuits:
We show that noisy quantum  circuits can simulate general
quantum circuits,  with an exponential cost.

\begin{ithm}\label{QUpper}
If a boolean function $f$ can be computed by a quantum circuit
of size $s$ and depth $d$, then $f$ can be computed by a 
noisy quantum circuit of size 
$O(s\cdot polylog(s))\cdot 2^{O(d\cdot polylog(d))}$ and depth
 $O(d\cdot polylog(d))$.
\end{ithm}

This gives a characterization of polynomial size noisy quantum
circuits.

\begin{corol}
Polynomial size noisy quantum circuits 
are not stronger than quantum circuits with $O(log(n))$ depth (The class 
$QNC^1$). On the other hand,
Quasi polynomial noisy quantum circuits can compute any function in $QNC^1$.
\end{corol}

For the lower bound we use the results in \cite{AB97},
in which, using noisy quantum circuits which allows the qubit to be 
initialized at different times, it is shown how to make the circuit
noise-resistant with polylogarithmic blow-up in the depth.
The reason for the fact  that in the quantum case the bounds are not tight,
is due to the fact that not as in classical circuits\cite{Von},
 it is yet unknown if quantum noise resistance can be achieved with constant 
blow-up in the depth. 

We emphasize again, that these results is
 very specific to the model we defined,
and as in the case of noisy reversible (classical) circuits,
 a slight change in the definitions changes dramatically the complexity 
power, and  variant definitions
of ``noisy quantum circuits'' 
turn out to be equivalent in power to normal quantum circuits,
due to the results in \cite{AB97,KLZ}.

\section{Noisy reversible circuits - The Upper Bound}
In this section we prove the upper bound, meaning that a noisy 
reversible circuit can simulate any circuit with exponential cost.
\begin{ithm}
If a boolean function $f$ can be computed by a boolean circuit
of size $s$ and depth $d$, then $f$ can be computed by a 
noisy reversible circuit of size $O(s \cdot 2^{O(d)})$ and depth $O(d)$.
\end{ithm}
{\bf Proof:} 
By \cite{Ben89}, we can convert the circuit that computes $f$ 
 to a reversible circuit, $R$, which has linear depth and polynomial size.
We now want to convert $R$ to $C$, a noise-resistant reversible circuit
 which computes $f$ with high probability.
 Note that the majority function can be implemented reversibility,
by a three bit to three bit gate, of which the first output is the majority 
of the three inputs.
Also, this reversible function on three bits can be implemented
 by a constant number of gates from the universal set being used, where these gates
will operate on the three bits plus a constant number of extra bits,
which will all be output in the state they where input.
To construct $C$, replace each bit in $R$ by $3^d$ bits.
Each time step we will limit the computation 
to a third of the bits, which will be ``good''.
In the $i'th$ time step we will operate on $3^{d-i}$ good bits.
This is done as follows:
A gate in the $i'$th time step in $R$, is transformed in $C$ to $3^{d-i}$
copies of the same gate, applied
 bitwise on the $3^{d-i}$ good bits.
We then divide these bits to triples, apply reversible majority gates on each triple, 
and limit ourselves to the  $3^{d-i-1}$ results of these gates, which 
will be the good bits, on which we operate bitwise the next time step in $R$,
and so on.
The claim is that if $p$ is small enough, the probability 
for ``good'' bit at time step $i$  to err is less than $p$.
The proof is by induction on $i$:
Let the input bits for the $i'$th step have error with probability  $\le p$.
 Another noise step makes this probability $\le 2p$.
After the computation step, if the fan-in of the gates is  $\le k$,
than the error probability for each output is $\le 2kp$.
After another noise step, the error probability is $\le(2k+1)p$. 
Now apply the majority gate.
Note that the error probabilities for each one of the inputs to the majority gates
 are independent.
Hence the error probability for the result of the majority gate 
is less than $3((2k+1)p)^2+((2k+1)p)^3$, which is  $\le p$
if $p$ is smaller than some constant threshold.\bbox

\section{Noisy Reversible Circuits - The Lower Bound}
In this section we prove the lower bound, meaning that 
after $O(log(n))$ steps of computation there is exponentially small
amount of information in the system.  
We first show that each step of faults, reduces the information in 
the system by a constant factor which depends only on the fault probability.
\begin{lemm}\label{exp}
Let $X$ be a string of $n$ bits, which is a random variable.
Let $Y$ be the string of $n$ bits generated from $X$ by 
 flipping each bit with independent 
probability $p$.
Then $I(Y)\le (1-2p)^2I(X)$, where $I$ is the Shannon information.
\end{lemm}
{\bf Proof:}
Let us first prove this for $n=1$.
Let $\alpha$,$\beta$ be the probability that the  bit $X,Y$ be $1$,
respectively. 
Let $\alpha=1/2+\delta/2$, 
\[ \beta=(1-p)\alpha+p(1-\alpha)=1/2+\delta \rho/2\]
where $\rho=1-2p,|\rho|\le 1$. Then $I(X), I(Y)$ are 
functions of $\delta$ and $\rho$. 
\[ I(X)=1+p_0log(p_0)+p_1log(p_1)=K(\delta)=((1+\delta)log(1+\delta)+(1-\delta)log(1-\delta))/2.\]
Developing $K(\delta)$ to a power series we get 
\[K(\delta)=(1/ln(2))\sum_{k=1}^{\infty}(\delta)^{2k}/[2k(2k-1)]\]
converging for all $0\le\delta\le1$. Therefore
\[I(Y)=K(\delta\rho)=
(1/ln(2))\sum_{k=1}^{\infty}(\delta\rho)^{2k}/[2k(2k-1)]\]
\[\le(1/ln(2))\rho^2\sum_{k=1}^{\infty}\delta^{2k}/[2k(2k-1)]=\rho^2K(\delta)=\rho^2I(X)\]
proving that 
\(I(Y)\le(1-2p)^2I(X).\)
We now use this result to prove for general $n$.
Let us write the strings $X,Y$ as $X_1,X_2,...,X_n$ and $Y_1,Y_2,...,Y_n$.
where $X_i$ and $Y_i$ are random variables that get the value $0$ or $1$.
\[I(Y)=\sum_{i=1}^{n} I(Y_i|Y_{i+1},..Y_n)\le 
\sum_{i=1}^{n} I(Y_i|X_{i+1},..X_n),\]
where we used the fact that 
\(I(A|C)\le I(A|B)\) where $A,B,C$ are random variables,
and  $C$  is a function 
of $B$, where the function might also be a random variable, independent of 
$A$ and $B$.
Using the following known formula:
  \(I(A|B)=\sum_{b} Pr(B=b)I(A|b)\), 
we can write the last term as
\[I(Y)\le \sum_{i=1}^{n} \sum_{x_{i+1},...x_n}
 Pr(X_{i+1}=x_{i+1}...X_n=x_n)I(Y_i|x_{i+1},..,x_n)
\le\]\[ \sum_{i=1}^{n} \sum_{x_{i+1},...x_n}
 Pr(X_{i+1}=x_{i+1}...X_n=x_n)(1-2p)^2I(X_i|x_{i+1},..,x_n)=
(1-2p)^2I(X),\]
where we have used the proof for one variable.
\bbox

We can now prove the main theorem:

{~}

\noindent{\bf Theorem \ref{Upper}:}
{\it For any noisy reversible circuit
of size $s$ and depth $d$ which is not worthless,
$s= 2^{\Omega(d)}$.}

{~}

{\bf Proof:}
We  first note that since each level of computation is reversible, 
the entropy does not change due to the computation step,
 and since the number of bits
is constant, the information does not change too during a computation
step.
We start with information $n$, and it reduces with rate which is exponential
in the number of noise steps:
After $m$ steps the information in the system is less than 
$(1-2p)^{2m}n$, by lemma \ref{exp}. When $m=O(log(n))$
 the information is polynomially small.
The information on any bit is smaller than
 the information on all the bits.\bbox

\section{Quantum Computation}
In this section we
 recall the definitions of quantum circuits\cite{Saq,Deu89,Yao93}
 with mixed states\cite{AN96},
 quantum noise, and quantum entropy
\cite{Per}.

\subsubsection{Pure states} 
 We deal with systems of  $n$ two-state quantum
particles, or ``qubits''. The {\em pure state} of such a system 
is a unit vector, denoted $|\alpha\rangle$,
in the Hilbert space\footnote{A Hilbert space is
 a vector space with 
an inner product} $\cal{C}$$^{2^{n}}$,
i.e. a $2^{n}$ dimensional complex space.
We view  $\cal{C}$$^{2^{n}}$ as a
 tensor product of $n$ two dimensional spaces, each corresponding to a qubit:
$\cal{C}$$^{2^{n}}= \cal{C}$$^{2}\otimes...\otimes\cal{C}$$^{2}$.
As a basis for   $\cal{C}$$^{{2}^{n}}$,
we use the $2^{n}$ orthogonal {\it basic states}:
 $|i\rangle=|i_{1}\rangle\otimes
|i_{2}\rangle....\otimes|i_{n}\rangle,0\le i< 2^{n}$,
 where $i$ is in binary representation,
and each $i_{j}$ gets 0 or 1.
Such a state corresponds to 
the j'th qubit being in the state 
   $|i_{j}\rangle$.
A pure state  $|\alpha\rangle\in\cal{C}$$^{{2}^{n}}$ 
is a {\em superposition}
 of the basic states:  
$|\alpha\rangle = \sum_{i=1}^{2^{n}} c_{i}|i\rangle$, 
 with $\sum_{i=1}^{2^{n}} |c_{i}|^{2}=1$. $|\alpha\rangle$ corresponds
to the vector 
 $v_{\alpha}=(c_{1},c_{2},...,c_{2^{n}})$.
 $v_{\alpha}^{\dagger}$, the complex conjugate of  $v_{\alpha}$,
is denoted  $\langle\alpha|$.
The inner product between $|\alpha\rangle$ and $|\beta\rangle$
is $\langle\alpha|\beta\rangle=
 (v_{\alpha},v^{\dagger}_{\beta})$.
The matrix  $v_{\alpha}^{\dagger}v_{\beta}$
 is denoted as  $|\alpha\rangle\langle\beta|$.
An isolated system of n qubits 
develops in time by a unitary matrix\footnote{Unitary
 matrices  preserve the
norm of any vector and satisfy the condition 
$U^{-1}=U^{\dagger}$}
 of size $2^{n} \times 2^{n}$:
 \( |\alpha(t_{2})\rangle = U|\alpha(t_{1})\rangle.\)
A quantum system in $\cal{C}$$^{{2}^{n}}$ can be {\em observed} by 
{\em measuring} the system.
An  important
 measurement is a {\it basic 
measurement} of
 a qubit $q$, of which the possible outcomes are $0,1$.
For the state  $|\alpha\rangle=\sum_{i=1}^{2^{n}} c_{i}|i\rangle$,
the  probability for outcome $0$ is $p_{0}= \sum_{i, i|_{q}=0}|c_{i}|^{2} $
and the state of the system 
will {\em collapse} to 
$|\beta\rangle=\frac{1}{p_{0}}\sum_{i, i|_{q}=0} c_{i}|i\rangle$,
 (the same for $1$).
In general,
an {\em observable} $O$ over $\cal{C}$$^{{2}^{n}}$  is  an
 hermitian\footnote{ An hermitian matrix $H$ satisfies $H=H^{\dagger}$} matrix, of size $2^{n}\times 2^{n}$. 
To apply a measurement of $ O$  on 
a pure state $|\alpha\rangle \in \cal{C}$$^{{2}^{n}}$,
write $|\alpha\rangle $  uniquely 
as a superposition
 of unit eigenvectors of $O$:
\(|\alpha\rangle= \sum_{i} c_{i}|o_{i}\rangle\),
where $|o_{i}\rangle$ have different eigenvalues. 
 With probability $ | c_{i}|^{2}$  the  measurement's outcome
 will be the eigenvalue of $|o_{i}\rangle$,
and the state  will {\it collapse} to $|o_{i}\rangle $.
A unitary operation $U$ on $k$ qubits  
 can be applied  on n qubits,
$n\geq k$, by taking the {\bf extension} $\tilde{U}$  of $U$,
i.e. the tensor product of $U$ with  an identity matrix on
 the other qubits.  The same applies for an observable $O$ to give
$\tilde{O}$.

\subsubsection{ Mixed states}
 A system which is not ideally isolated  from 
it's environment is described by a {\em mixed state}.
There are two equivalent descriptions of mixed states:
mixtures and density matrices.
We use density matrices in this paper.
A system in the {\bf mixture}  $\{\alpha\}=\{p_{k},|\alpha_{k}\rangle\}$
is  with probability $p_{k}$
in the pure state $|\alpha_{k}\rangle$.
The rules of development in time and measurements 
for mixtures are 
obtained by applying {\bf classical} probability to
the rules 
for pure states.
A density matrix $\rho$ 
on $\cal{C}$$^{2^{n}}$ is an hermitian positive semi definite complex matrix
 of dimentions $2^{n}\times 2^{n}$,
with $tr(\rho)=1$.
A pure state $|\alpha\rangle=\sum_{i} c_{i}|i\rangle$ 
is associated the density matrix
 \(\rho_{|\alpha\rangle} = |\alpha\rangle\langle\alpha|\) i.e.  
\(\rho_{|\alpha\rangle}(i,j)= c_{i}c_{j}^{*}.\)
A mixture 
$\{\alpha\}=\{p_{l},|\alpha_{l}\rangle\}$,
is  associated the density matrix :
\(\rho_{\{\alpha\}} = \sum_{l} p_{l} \rho_{|\alpha_{l}\rangle}.\)
The operations on a density
matrix are defined such that  the correspondence to mixtures is preserved.
If a unitary matrix $U$ transforms the mixture 
\(\{\alpha\}=\{p_{l},|\alpha_{l}\rangle\}\) to
\(\{\beta\}=\{p_{l},U|\alpha_{l}\rangle\},\)
then
 \(\rho_{\{\beta\}} = \sum_{l}  p_{l}
 U|\alpha_{l}\rangle\langle\alpha_{l}|U^{\dagger}=
U\rho_{\{\alpha\}}U^{\dagger}.\)
Let  $\rho$ be written in a basis of
 eigenvectors $v_{i}$ of an observable $O$.
A measurement of $O$ on $\rho$
gives, the outcome $\lambda$ with the probability which is
 the sum of the diagonal terms of $\rho$, which relate to 
the eigenvalue $ \lambda$: 
$pr(\lambda)=\sum_{i=1}^{2^{n}} \rho_{v_{i},v_{i}} \delta(\lambda_{i}
=\lambda)$.
conditioned that the outcome is the eigenvalue $\lambda$,
the resulting density matrix is $O_{\lambda}\circ(\rho)$, which we get
by  first
putting to zero all rows and columns in $\rho$, which relate 
to eigenvalues different from
$\lambda$, and then renormalizing this matrix to trace one. 
Without conditioning  on the outcome 
the resulting density matrix will be  
\(O\circ(\rho)=\sum_{k}
 Pr(\lambda_{k}) O_{\lambda_{k}}\circ(\rho). \)
which differs from  $\rho$, only in
 that the entries in  $\rho$ which connected between
 different eigenvalues are put to zero. 
Given a density matrix $\rho$ of n qubits,
 the reduced density matrix of a subsystem,$ A$,
 of, say, $m$ qubits is defined as an average over the states of
the other qubits:   
 \( \rho|_{A}(i,j)= \sum_{k=1}^{2^{n-m}} \rho(ik,jk)\).

\subsection{\bf Quantum circuits with mixed states}
{\em A quantum unitary gate} of order $k$ is a complex unitary 
matrix of size $2^{k} \times 2^{k}$. 
A density matrix $\rho$ will transform
by the gate to  \(g\circ\rho = \tilde{U}\rho\tilde{U}^{\dagger}\),
where $\tilde{U}$ is the extension of $U$.
{\em A measurement gate} of order $k$ is a complex hermitian 
matrix of size $2^{k} \times 2^{k}$.
A density matrix $\rho$ will transform
by the gate to  $g\circ\rho = \tilde{O}\circ(\rho)$. 
{\em A Quantum circuit} 
 is a directed acyclic graph
with $n$ inputs and $n$ outputs. 
 Each node $v$ in the graph is labeled by a quantum gate $g_{v}$.
The in-degree and out-degree of $v$ are equal to the order of $g_{v}$.
Some of the outputs are labeled ``result'' to indicate that
these are the qubits that will give the output of the circuit.
The wires in the circuit correspond to qubits.
An initial density matrix $\rho$ 
transforms by a circuit $Q$ to a 
 final density matrix  $Q\circ \rho =
g_{t}\circ...\circ g_{2}\circ g_{1}\circ \rho$,
where the gates  $ g_{t}...g_{1}$ are applied in a topological order. 
For an input string $i$,
the initial density matrix is $\rho_{|i\rangle}$.
The output of the circuit is the outcome of applying basic measurements
of the result qubits, on the final density matrix  
$Q\circ\rho_{|i\rangle}$. Since the outcomes of measurements
 are random, the function that the circuit computes is a 
{\it probabilistic function}, i.e. for input $i$ it outputs 
strings according to a distribution which depends on $i$.

\subsection{Noisy Quantum Circuits}
As any physical system, a quantum system is subjected to noise.
The process of errors depends on time, so the
 quantum circuit will be divided to levels, or time steps.
In this model, (as in \cite{Yao93} but not as in \cite{AB97})
 all qubits are present at time $0$.
The model of noise we use for noisy quantum circuits is a single qubit noise,
in which a qubit is damaged with probability $1/2>p>0$ each time step.
The damage operates as follows: 
A unitary operation operates on the qubit and a state of
 the environment (The environment can be represented by $m$
 qubits in some state).
This operation results in a density matrix of the $n$ qubits of the system
and the environment. We reduce this density matrix to the $n$ qubits of the 
circuit to get the new density matrix after the damage.
The density matrix of the circuit develops by applying alternately the 
computation step and this probabilistic process of noise.
The function computed by the noisy quantum circuit is naturally 
the average over the outputs, on  the probabilistic process of noise.

\subsection{Quantum Entropy}
In this subsection we give some background about the notion of quantum 
entropy and it's relation to Shannon information.
All deffinitions and lemmas can be found in the book of Asher Peres\cite{Per}.

\begin{deff}\label{ent1}
The (von Neumann) entropy of a density matrix $\rho$ is defined to be
\(S(\rho)= -Tr(\rho log_{2}(\rho))\).
\end{deff}

\begin{deff}\label{ent2}
The information in a density matrix $\rho$ of $n$ qubits is defined   
to be $I(\rho)=n-S(\rho)$.
\end{deff}

The Shannon
 entropy, $H$, in the distribution over the  
results of any measurement on $\rho$ is larger then
the Von-Neumann entropy in $\rho$.
This means that one can not extract more Shannon
 information from $\rho$ than 
the Von Neumann information.
\begin{lemm}\label{ent3}
Let $O$ be an observable of $n$ qubits, $\rho$ a density matrix of $n$
qubits.
Let $f$ be the distribution which $\rho$ induces on the eigenvalues of $O$.
Then \(H(f)\ge S(\rho)\).
\end{lemm}
As the Shannon entropy, the Von-Neumann entropy  is concave:
\begin{lemm}\label{ent4}
Let $\rho_i$ be density matrices of the same number of qubits.
Let $p_i$ be some distribution on these matrices.
\(S(\sum_i p_i\rho_i)\ge \sum_i p_i S(\rho_i).\)
\end{lemm}
The Shannon entropy of two independent variables is just the 
sum of the entropies of each one.
The Quantum analog is that the Von-Neumann entropy
 in a system which consists of 
non-entangled subsystems 
is just the sum of the entropies:
\begin{lemm}\label{ent5}
\(S(\rho_1)+S(\rho_2)=S(\rho_1\otimes\rho_2)\).
\end{lemm}
One can define a relative Von-Neumann entropy:  
\begin{deff}\label{ent6}
Let $\rho_1,\rho_2$ be two density matrices of the same number of qubits.
The {\bf relative entropy} of $\rho_1$ with respect to $\rho_2$ is defined as 
\(S(\rho_1 |\rho_2)= Tr[\rho_2(log_{2}(\rho_2)-log_{2}(\rho_1))].\)
\end{deff}
The relative entropy is a non-negative quantity:
\begin{lemm}\label{ent7}
Let $\rho_1,\rho_2$ be two density matrices of the same number of qubits.
The relative entropy of $\rho_1$ with respect to $\rho_2$
 is non negative: \(S(\rho_1 |\rho_2)\ge 0\).
\end{lemm}

\section{The Quantum upper Bound}
In this section we prove the upper bound for the case of noisy 
quantum circuits:  a noisy 
quantum circuit can simulate any quantum circuit with exponential cost.

{~}

\noindent{\bf Theorem \ref{QUpper}:}
If a boolean function $f$ can be computed by a quantum circuit
of size $s$ and depth $d$, then $f$ can be computed by a 
noisy quantum circuit of size 
$O(s\cdot polylog(s))\cdot 2^{O(d\cdot polylog(d))}$ and depth
 $O(d\cdot polylog(d))$.

{~}

{\bf Proof:} 
In \cite{AB97} it is shown that any quantum circuit $Q$, with depth $d$
and size $s$, can be simulated by a  noisy quantum circuit $\tilde{Q}$, 
 with depth polylogarithmic in $d$ and size  polylogarithmic in $s$,
 where a different model
of noisy quantum circuit is used: qubits are allowed to 
be initialized at any time during the computation. 
To adapt the proof to our model, in which all qubits are present at time $0$,
it will suffice to show that there exists a 
noisy quantum circuit, $A_t$, with depth $t$ operating on $3^{t}$ qubits,
such that if the error probability is $p$, for an input string of all zeroes,
at time $t$  the first qubit is in the state $|0>$
with probability  $\ge (1-p)$.
If such $A_t$ exists, than for  each qubit, $q$,
  which is input to $\tilde{Q}$ at time $t$, we simply add $3^{t}-1$
 qubits to the circuit, and together with $q$
they will be initialized at $t=0$ to be $|0>$. 
On these $3^{t}$ qubits we will operate the sequence of gates $A_t$,
and the first qubit will play the role of $q$ after time $t$.
The new circuit is a noisy quantum circuit for which all
 qubits are initialized  at time $0$, it's size is
 $O(s\cdot polylog(s))\cdot 2^{O(d\cdot polylog(d))}$ and depth
 $O(d\cdot polylog(d))$.

 $A_t$ is constructed as follows:
We begin with $3^{t}$ qubits
in the state $|0>$. These qubits can be divided to triples of qubits.
We apply the following ``majority'' quantum gate,  on each  triple:

\begin{quote}
$|000> \longmapsto |000>~~,~~|100> \longmapsto |011>$

$|001>\longmapsto  |001>~~,~~|101>  \longmapsto |101>$

$|010> \longmapsto |010>~~,~~|110>  \longmapsto |110>$

$|011> \longmapsto |100>~~,~~|111> \longmapsto |111>.$
\end{quote}
The first qubit of each triple carries now the result of the majority.
(Note that the function of majority works here, in the quantum case, because
we only need to deal with non-entangled states: the zero $|0>$ state.
The majority gate is not a good method to pick the majority out of three
general pure states.)
All these $3^{t-1}$ result
 qubits can now be divided also to triples,  and we apply  
majority gates on these triples, and so on, until time $t$.
We claim  that the error probability of the
 result qubits of time step $i\le t$ 
 is  $\le p$, if $p$ is smaller than some threshold.
Let us prove this by induction on $i$.
If each qubit in the $i$'th time step has
 error probability $\le p$, than after one noise step it's error probability
is $\le 2p$.
The majority gate is applied on qubits with independent error probabilities.
Thus the probability for the majority result to  err
is less than $3(2p)^2+(2p)^3$.
This probability is smaller than $p$ if $p$ is small enough. \bbox

\section{The Quantum Lower Bound}  
We prove that  in a noisy quantum circuit
 the information decreases exponentially
in the number of time steps, in the presence of the
  following type of quantum noise, where a qubit 
that undergoes a fault   is replaced by a qubit in one of the basic states,
which is chosen randomly. Such a qubit carries no information.
We first show that a noise step causes the information in the circuit
to decrease at least by a constant factor $(1-p)$, and then show that 
quantum gates can only decrease the information in the system.
This is true even for the more general model of quantum non-reversible
circuits, where measurement gates are allowed.

Let us first show that quantum gates can only reduce information:
\begin{lemm}\label{Qgate}
Let $g$ be a quantum gate, $\rho$ a density matrix.
$I(g\circ \rho)\le I(\rho)$.
\end{lemm}

{\bf Proof:} For a unitary gate, which is reversible,
  $I(\rho)= I(g\circ\rho)$.
The proof for the case of a measurement gate is given in the appendix.\bbox

In order to show that during a noise step the information
 decreases by a factor, we need the following lemma, 
(which is the quantum analog of a theorem proved in \cite{}): 
The average information in $k$ qubits chosen randomly
out of $n$ qubits, is smaller then 
$\frac{k}{n}$  the information in all the $n$ qubits.

\begin{lemm}\label{average}
Let $\rho$ be a density matrix of $n$ qubits,
and let $k<n$.
Then \(\frac{1}{\left(\begin{array}{c}n\\k\end{array}\right)}\sum_{A_{k}}I(\rho|_{A_{k}})\le \frac{k}{n}I(\rho).\)
\end{lemm}

We can now prove that for a specific type of quantum noise, that in which 
 qubit is replaced with probability $p$ by a qubit in a random state,
 the information in the quantum circuit decreases
by a factor of $(1-p)$ after each noise step:

\begin{lemm}\label{Qdec}
Let $\rho$ be a density matrix of $n$ qubits.
Let each qubit in $\rho$ be replaced with independent probability
$p$ by a qubit in the density matrix $\rho_{R}=\frac{1}{2}(|0\rangle\langle 0|+
|1\rangle\langle 1|)$, to give the density matrix $\sigma$.
Then $I(\sigma)\le (1-p)I(\rho)$.
\end{lemm}
{\bf Proof:}

Let us write $\sigma=\sum_{k=1}^{n}\sum_{A_{k}}p^{n-k}(1-p)^k \rho|_{A_{k}}\otimes \rho_{R}^{n-k}$,
where the sum over $A_{k}$ is a sum over all subsets of $k$ qubits,
and the power on the density matrices means taking $n-k$ times the {\it 
tensor product} of $ \rho_{R}$.
This presents the resulting density matrix as a probability distribution 
over all possible cases where the faults could have occured, with the correct
probabilities.
By the concavity of the entropy we have:
\(I(\sigma)\le \sum_{k=1}^{n}p^{n-k}(1-p)^k \sum_{A_{k}}[I(\rho|_{A_{k}})+
(n-k) I(\rho_{R})]\),
where we have used lemma \ref{ent5}.
Since $I(\rho_{R})=0$, we have that 
\begin{equation}\label{help}
I(\sigma)\le\sum_{k=0}^{n}p^{n-k}(1-p)^k \sum_{A_{k}}I(\rho|_{A_{k}}).
\end{equation}
Using lemma \ref{average}, we get 
\(I(\sigma)\le\sum_{k=1}^{n}p^{n-k}(1-p)^k \frac{k}{n}\left(\begin{array}{c}n\\k\end{array}\right)I(\rho)=(1-p)I(\rho)\).\bbox

We can now prove the lower bound on noisy quantum circuits:

{~}

\noindent{\bf Theorem \ref{QLower}:}
{\it For any noisy quantum circuit
of size $s$ and depth $d$ which is not worthless,
$s= 2^{\Omega(d)}$.}

{~}

{\bf Proof:}
Using lemmas \ref{Qgate} and \ref{Qdec}
 we can show by induction on $t$ that after $t$ time steps,
 the information in the system $I(\rho)\le(1-p)^ts$, so the information in
the final density matrix is $\le(1-p)^ds$. 
The classical information in the probability distribution
which we get when measuring the result qubits in $\rho$ is smaller than
 $I(\rho)$, due to lemma \ref{ent3} and
 the fact that the  basic measurements of the $r$ result
qubits can be replaced  by one observable on  the result qubits,
with each possible string $|i\rangle$ as an eigenvector with eigenvalue $i$.
\bbox
\section{Open Question}
We have shown that the power of noisy reversible circuit
is as the complexity class $NC^1$.
Is the power of noisy quantum circuit exactly that of 
the quantum analog complexity class, $QNC^1$?
Making the lower bound tight connects to the following open question:
Can noisy quantum circuits be made noise resistance with only 
a constant blow-up in depth?  
\bibliographystyle{plain}
\bibliography{/CS/grad/doria/references}

\section{Appendix-Quantum Entropy lemmas}
In this appendix we give the proofs of lemmas regarding quantum entropy.
All the proofs, except the last one which is a new result as far as we know,
are taken from \cite{Per}.

{~}

{\bf Lemma \ref{ent3}:}
Let $O$ be an observable of $n$ qubits, $\rho$ a density matrix of $n$
qubits.
Let $f$ be the distribution which $\rho$ induces on the eigenvalues of $O$.
Then \(H(f)\ge S(\rho)\).

{~}

{\bf Proof:}
Let $\{v_{i}\}$ be the eigenvectors of $\rho$, with $P_i$ there eigenvalues.
$S(\rho)=-\sum_i P_i log(p_i)$.
The probability to get an eigenvalue $\lambda_j$ measuring $\rho$ is
 $Q_j=\sum_{k}P_kG_{k,j}$, where $G_{k,j}$ is the probability to get 
 $\lambda_j$ when measuring $v_{k}$.
So  $S(f)=-\sum_j Q_j log(Q_j)$.
\(S(f)-S(\rho)=\sum_i P_i log(p_i)-\sum_j Q_j log(Q_j)=
\sum_i  P_i(log(p_i)-\sum_{j}G_{i,j} log(Q_j))=
\sum_{i,j}  P_iG_{i,j} log(P_i/Q_j))\).
Where we have used $\sum_{j}G_{i,j}=1$.
Now $logx\ge 1-\frac{1}{x}$,
to give that  
\(S(f)-S(\rho)\ge \sum_{i,j}  P_iG_{i,j}(1- Q_i/P_j)=0\).\bbox

{~}

{\bf Lemma \ref{ent4}:}
Let $\rho^i$ be density matrices of the same number of qubits.
Let $p^i$ be some distribution on these matrices.
\(S(\sum_i p^i\rho^i)\ge \sum_i p^i S(\rho^i).\)

{~}

{\bf Proof:}
Let us interprate the diagonal terms in a density matrix $\rho$
as a classical  probability distribution $D$. We have that 
in any basis, $H(D)\le S(\rho)$, where the equality is achieved if and only if
 the $\rho$ is written in the basis which diagonalized it.
Let us write all matrices in the basis which diagonalizes the matrix 
$\rho=\sum_i p^i\rho^i$,
 and let the diagonal terms of
$\rho$ be the distribution $D$ and the diagonal terms in $\rho_i$ be
the distributions $D_i$, respectively.
We than have
$\S(\rho)=H(D)=H(\sum_i p_i D_i),$
where the sum is in each coordinate, and by the concavity 
 of classical entropy, we have that $H(\sum_i p_i D_i)\le
 \sum_i p_i H(D_i)$,
but since $H(D_i)\le S(\rho_i)$ it closes the proof.\bbox.

{~}

{\bf Lemma \ref{ent5}:}
\(S(\rho_1)+S(\rho_2)=S(\rho_1\otimes\rho_2)\).

{~}

{\bf Proof:}
If $\{\lambda^1_i\},\{\lambda^2_j\}$ are the sets of eigenvalues of
$\rho_1,\rho_2$ respectively,
the eigenvalues of $\rho_1\otimes\rho_2$ are just
$\{\lambda^1_i\lambda^2_j\}$,
and the entropy is  
\(S(\rho_1\otimes\rho_2)=-\sum_{i,j}\lambda^1_i\lambda^2_j log(\lambda^1_i\lambda^2_j)=S(\rho_1)+S(\rho_2)\).\bbox

{~}

{\bf Lemma \ref{ent7}:}
Let $\rho_1,\rho_2$ be two density matrices of the same number of qubits.
The relative entropy of $\rho_1$ with respect to $\rho_2$
 is non negative: \(S(\rho_1 |\rho_2)\ge 0\).

{~}

{\bf Proof:}
Let $\{|v^1_{m}\rangle\},\{|v^2_{m}\rangle\}$ be the eigenvectors of 
$\rho_1$,$\rho_2$
respectively, and $\{\lambda^1_{m}\},\{\lambda^2_{m}\}$ be the corresponding
eigenvalues, respectively.
We can write $\log(\rho_2)=\sum_{m}log(\lambda^2_m)
|v^2_{m}\rangle\langle v^2_{m}|$.
We want to evaluate the relative entropy 
\(S(\rho_1 |\rho_2)= Tr[\rho_2(log_{2}(\rho_2)-log_{2}(\rho_1))]\)
in the first basis $\{|v^1_{m}\rangle\}$ where $\rho_1$ is diagonal.
The diagonal elements of $log(\rho_2)$ in this basis are:
\(log(\rho_2)_{m,m}=\sum_{n}log(\lambda^2_n)
\langle v^1_{m}|v^2_{n}\rangle\langle v^2_{n}| v^1_{m}\rangle=
\sum_{n}log(\lambda^2_n)|\langle v^1_{m}|v^2_{n}\rangle|^{2}\),
so  
\(S(\rho_1 |\rho_2)=\sum_{m}\lambda^1_m(log(\lambda^1_m)-
\sum_{n}log(\lambda^2_n)|\langle v^1_{m}|v^2_{n}\rangle|^{2})=
\sum_{m,n}\lambda^1_m  |\langle v^1_{m}|v^2_{n}\rangle|^{2} 
log(\frac{\lambda^1_m}{\lambda^2_n}),\)
where we have used $\sum_{m}|\langle v^1_{m}|v^2_{n}\rangle|^{2}=1$.
Since $log(x)\ge 1-\frac{1}{x}$,  
\(S(\rho_1 |\rho_2)\ge \sum_{m,n}
\lambda^1_m  |\langle v^1_{m}|v^2_{n}\rangle|^{2} 
(1-\frac{\lambda^2_n}{\lambda^1_m})=
\sum_{m}\lambda^1_m\sum_{n}|\langle v^1_{m}|v^2_{n}\rangle|^{2}-
 \sum_{n}\lambda^1_n\sum_{m}|\langle v^1_{n}|v^2_{m}\rangle|^{2}= 0.\)\bbox

Let us prove another fact which will be needed.

\begin{lemm}\label{ent8}
Let $\rho_1$ be a reduced density matrix  of $\rho$ to a subsystem A. 
Then
$-Tr(\rho log(\rho_1\otimes I))=-Tr(\rho_1 log(\rho_1))=S(\rho_1)$.
\end{lemm}
{\bf Proof:}
Let us write everything in a basis for the whole system,
 where $\rho_1$ is diagonal.
 In this basis $log(\rho_1\otimes I))_{in,in}=
log((\rho_1)_{i,i}$,
where the first index (i or j) indicates qubits in $A$, and the second index
indicates the qubits which we disregard and trace over.
$-Tr(\rho log(\rho_1\otimes I))=-\sum_{i,n}( \rho_{in,in}log((\rho_1)_{i,i})=
-\sum_{i}( \rho_{i,i}log((\rho_1)_{i,i})=S(\rho_1)$,
 where we have used the fact that 
 $\rho_1$ is a reduced density matrix which satisfies
$(\rho_1)_{i,j}=\sum_{n}(\rho)_{in,jn}$.\bbox

{~}

\noindent{\bf Lemma \ref{Qgate}:}
Let $g$ be a quantum gate, $\rho$ a density matrix.
$I(g\circ\rho)\le I(\rho)$.

{~}

{\bf Proof:}
The entropy is invariant under unitary transformation, since unitary 
transformations change the eigenvectors, but does not change the
 set of eigenvalues which is what 
determines the entropy of the density matrix.
Therefore the information does not change if $g$ is unitary.
If $g$ is a measurement gate,
 let us write $\rho,g\circ\rho$ in
a basis of eigenvectors of the extension of the observable $g$,
in which $ g\circ\rho$ is diagonalizes.
By lemma \ref{ent7}, the relative entropy of $g\circ\rho$ with respect to
 $\rho$ is non negative. Writing the relative entropy in the basis of 
eigenvectors: 
$0\le S(g\circ\rho|\rho)=
Tr[\rho(log_{2}(\rho)-log_{2}(g\circ\rho))]=
-S(\rho)-  \sum_{m}\rho_{m,m}log((g\circ\rho)_{m,m})=-S(\rho)+S(g\circ\rho)$,
where the last equality is due to the fact that in this basis 
$\rho_{m,m}=(g\circ\rho)_{m,m}$. Hence $n-S(g\circ\rho)\le n-S(\rho)$.\bbox

{~}

\noindent{\bf Lemma \ref{average}:}
Let $\rho$ be a density matrix of $n$ qubits,
and let $k<n$.
Then \(\frac{1}{\left(\begin{array}{c}n\\k\end{array}\right)}\sum_{A_{k}}I(\rho|_{A_{k}})\le \frac{k}{n}I(\rho).\)

{~}

{\bf Proof:} Let $\rho_{1}$  be the
 tensor product of $k\left(\begin{array}{c}n\\k\end{array}\right)$ 
copies of $\rho$,
and let $\rho_{2}$ be the tensor product of all the possible reduced 
$\rho|_{A_{k}}$, each taken $n$ copies.
$\rho_{1}$ and $\rho_{2}$ are matrices of an equal number of qubits:
  $nk\left(\begin{array}{c}n\\k\end{array}\right)$.
Hence we can use the non negativity of the relative entropy
(  lemma \ref{ent7}),
 and
write \(0\le S(\rho_{2}|\rho_{1})= Tr(\rho_1(log(\rho_1)-log(\rho_2)))=
-S(\rho_1)-Tr(\rho_1 log(\Pi_{A_{k}}(\rho|_{A_{k}})^{n}))=
- k\left(\begin{array}{c}n\\k\end{array}\right) S(\rho)-
\sum_{A_{k}}Tr(\rho_1log((\rho|_{A_{k}})^n\otimes I^{n})),\)
where all products and powers of matrices are understood as tensor products,
and using  the fact that the logarithm of tensor products can be written
as the sum of logarithms.
We now observe that  $(\rho|_{A_{k}})^{n}$
is a reduced density matrix  of $\rho_1$ if $k>0$.
 Lemmas \ref{ent8} and \ref{ent5} imply that 
\(0\le- k\left(\begin{array}{c}n\\k\end{array}\right) S(\rho)+
\sum_{A_{k}} nS(\rho|_{A_{k}})\),
so for $k>0$, 
$\sum_{A_{k}}I(\rho|_{A_{k}})\le 
\frac{k}{n}\left(\begin{array}{c}n\\k\end{array}\right)I(\rho).$\bbox

\end{document}